\documentstyle[12pt,axodraw]{article}
\newcommand{\beq}{\begin{eqnarray}}   
\newcommand{\eeq}{\end{eqnarray}}
\newcommand{\ra}{\rightarrow}

\newcommand{\gsim}{\lower.7ex\hbox{$
\;\stackrel{\textstyle>}{\sim}\;$}}
\newcommand{\lsim}{\lower.7ex\hbox{$
\;\stackrel{\textstyle<}{\sim}\;$}}

\begin{document}
\begin{titlepage}
\renewcommand{\thefootnote}{\fnsymbol{footnote}}

\begin{center} \Large
{\bf Theoretical Physics Institute}\\
{\bf University of Minnesota}
\end{center}
\begin{flushright}
TPI-MINN-97/14-T\\
UMN-TH-1543-97\\
hep-ph/9705417\\
\end{flushright}
\vspace*{1cm}

\begin{center}
{\Large \bf  Exact Results for Soft Supersymmetry Breaking 
Parameters in Supersymmetric Gauge  Theories}
\vspace{0.8cm}

{\Large J. Hisano}\footnote
{Address after  September 1, 1997: 
National Laboratory For High Energy Physics (KEK),
1-1 Oho, Tsukuba-shi, Ibaraki-ken 305, Japan.
} 

\vspace{0.3cm}
{\it and}

\vspace{0.3cm}

{\Large  M. Shifman} 

\vspace{0.5cm}
{\it  Theoretical Physics Institute, Univ. of Minnesota,
Minneapolis, MN 55455}

\end{center}

\vspace*{.3cm}

\begin{abstract}

We obtain exact relations (valid to all orders in the coupling 
constant)
for the running gaugino mass in supersymmetric gauge theories, 
treating the soft supersymmetry breaking effects in the linear 
approximation. If a supersymmetry breaking squark (selectron) 
mass term is
introduced, our relation connects the renormalization group equation 
for this term with that for the gaugino mass. 
Exact relations for the threshold effects in the gaugino masses
are derived. 
The key ingredients of
our analysis is the use of the Wilsonean action and holomorphy 
of this action with respect to relevant parameters.

\end{abstract}

\end{titlepage}

\section{Introduction}

Holomorphy is one of the most powerful tools in explorations
of supersymmetric (SUSY) gauge theories.  Historically,
the first exact result, the so called NSVZ $\beta$ function
was obtained \cite{NSVZ} by exploiting the holomorphy of the
gauge kinetic term in the Wilsonean action (for a recent review 
see Ref.~\cite{RR}). Later on new insights 
were obtained from similar ideas in a
wide range of theories with superpotentials \cite{Seiberg}.
The role of the holomorphic anomaly was revealed
\cite{Shif}.

In this paper we report a new, so far unexplored  application of the 
method
based on holomorphy in SUSY gauge theories with {\em softly 
broken} supersymmetry.  The SUSY breaking parameters --
the gaugino mass $m_{\tilde g}$ and the squark (selectron) mass 
term 
$m_{\tilde q}$ -- are considered in
the linear approximation (i.e.  we disregard effects containing
powers of $m_{\tilde g, \tilde q}$ higher than the first), but to all
orders in the coupling constant. We obtain the renormalization group 
(RG) equations governing the running of these parameters,
valid {\em to all
orders} in the coupling constant. The simplest example of this type 
emerges in supersymmetric gluodynamics, i.e. the theory of gluons 
and gluinos, without matter fields. In this model
the combination
\beq
\frac{\alpha m_{\tilde g}}{\beta (\alpha )} = \mbox{RGI}\, ,
\label{RGI}
\eeq
where RGI stands for RG invariant, and $\beta$ is the Gell-Mann-Low
function,
\beq
\beta (\alpha ) =-\frac{\alpha^2}{2\pi }\,\, \frac{3T(G)}{1-
(T(G)\alpha
/2\pi)}
\, ,
\label{GMLF}
\eeq
where $T(G)$ is (one-half) of the Dynkin index ($T(G)=N$ for 
$SU(N)$ theories).  In the one-loop approximation the left-hand side 
of Eq.~(\ref{RGI}) reduces to $m_{\tilde g}/\alpha$. The fact that
this ratio is RG invariant in the leading approximation is well-known.
Equation~(\ref{RGI}) generalizes this result to all orders. The formula 
(\ref{RGI}) is general. It holds also in  supersymmetric gauge 
theories with matter (with the corresponding $\beta$ function, see 
Eq. (\ref{NSVZbeta})) provided there are no super-Yukawa (trilinear) 
couplings in the superpotential. If trilinear couplings are introduced, 
Eq. (\ref{RGI})
is replaced by a more general relation, see Eq. (\ref{mb2}) below. 

A particular but very interesting issue belonging to the given range 
of questions is the impact of the mass thresholds. We show how the 
threshold effects can be exactly incorporated in the gluino mass.

Section 2 introduces our notation and conventions. 
We begin our analysis (Sect. 3) with a simple case of supersymmetric 
electrodynamics (SQED). In this problem non-trivial dynamics arises
only from loops with the matter fields. One can consider the soft 
supersymmetry breaking due to the photino mass and due to the 
selectron mass term. The later will be chosen in a special form.
We then derive an exact RG relation between these parameters.

In Sect. 4 softly broken non-Abelian gauge theories are considered.
A subtle issue is to which particular action the holomorphy-based 
arguments apply. Two distinct constructions go under the name 
``effective action":  the first, $\Gamma (\mu )$, is the generator
of one-particle irreducible vertices, the second, $S (\mu )$,
is the Wilsonean effective action, where all infrared contributions
are excluded, by definition. As was shown in Ref.~\cite{ShiVa}
the holomorphic dependence refers to the Wilsonean action.
At the same time, such parameters as the gauge coupling constants 
and the gluino mass are introduced through $\Gamma$. Exact 
results for the renormalization of the gluino mass can be obtained
due to the fact that the relation between the parameters in $ 
\Gamma$ and $S$ is known. All parameters appearing in the 
Wilsonean action will be marked by the subscript W.

We also consider in Sect. 5 a toy model of Grand Unification (GUT)
and derive prototype ``GUT relations" valid to all orders.
In Sect. 6 we confront  all-order predictions with explicit 
two-loop calculations of the gluino mass known in the literature,
and find perfect agreement. Finally, Sect. 7 summarizes our results.

\section{Preliminaries}

In this section we will briefly review our notations and conventions 
and discuss a mechanism through which the soft SUSY breaking
parameters will be introduced. 

Supersymmetric generalization of pure gluodynamics,  the theory of 
gluons and gluinos, is described by the component Lagrangian  
\cite{FZ}
\beq
{\cal L}_{\rm SYM} =  -\frac{1}{4g_0^2} 
G_{\mu\nu}^aG_{\mu\nu}^a + {\vartheta\over 
32\pi^2}G_{\mu\nu}^a\tilde G_{\mu\nu}^a
+\frac{1}{g_0^2}\left[
i\lambda^{a \alpha} 
D_{\alpha\dot\beta}\bar\lambda^{a\dot\beta}
\right] \, 
,
\label{SUSYML}
\eeq
where the spinorial notation is used. In the superfield language
the Lagrangian can be written as
\beq
{\cal L}_{\rm SYM} =  \int d^2\theta \frac{1}{4g^2} \mbox{Tr} W^2 + 
{\rm H.c.}\, ,
\label{SFYML}
\eeq
where
$$
\frac{1}{g^2} = \frac{1}{g_0^2} -\frac{i\vartheta}{8\pi^2}\, .
$$
In what follows the vacuum angle $\vartheta$ will play no
special  role.
It is important, however,  that $g^{-2}$ can be treated as a complex
parameter.
  Our conventions regarding the superfield formalism 
are summarized e.g. in the recent review
\cite{Shif2}
\footnote{These conventions are essentially those of Bagger and 
Wess \cite{BW} .
The distinctions are that we use the metric $(+---)$ and
the Grassmannian differentials are normalized as 
$\theta^2 d^2\theta  = 2$.}. We will limit ourselves to the $SU(N)$ 
gauge 
group
(the generators of the group $T^a$ are in the fundamental 
representation, so that $\mbox{Tr}(T^aT^b) = (1/2)\delta^{ab}$).

The matter fields are assumed, for simplicity, to belong to the 
fundamental
representation of $SU(N)$. (Our final results are independent of this 
assumption.) In this case each flavor consists of two subflavors,
$Q$ and $\tilde Q$. These  superfields are  in the representation $N$
and $\overline{N}$ of the gauge group, respectively.
The Lagrangian of the matter sector has the form
\beq
{\cal L}_{\rm M} = \frac{1}{4}\int d^2\theta d^2\bar\theta^2
\left( \bar Q  e^V Q + \bar{\tilde Q}  e^{-V} \tilde Q \right) +\left( 
\int d^2\theta \frac{m_0}{2} 
Q^{\alpha }\tilde Q_{\alpha } +\mbox{H.c.}\right)\, ,
\label{matter}
\eeq
where $\alpha$ is the color  index, $\alpha  = 
1,2, ..., N$.   The subscript
0 of the matter mass term $m_0$  indicates that it is the bare mass  
that enters the original Lagrangian;
this parameter is complex. 
It is assumed that the matter mass matrix is diagonal in flavor.
Such a diagonalization can always be achieved.

In SQED the gauge part of the Lagrangian takes the form
\beq
{\cal L}_{\rm SQED} =  \int d^2\theta \frac{1}{8g^2} W^2 + {\rm H.c.}
\label{sqedl}
\eeq
while the matter part is the same as in Eq.~(\ref{matter}) with the
omission of the color indices.

Now we must discuss how the soft supersymmetry breaking is 
introduced. To this end $1/g^2$ in Eqs.~(\ref{SFYML}) or
(\ref{sqedl}) is substituted by a chiral superfield
${\cal S}$, so that the expectation value of the lowest component
$$
\langle {\cal S} \rangle = \frac{1}{g^2}\, .
$$
The expectation value of the $F$ component generates
the gluino (photino) mass $m_{\tilde g}$, namely
\beq
{\cal S} \ra \frac{1 - 2 m_{\tilde g}\theta^2}{g^2}\, .
\label{gmg}
\eeq

By the same token we substitute the parameter $m$ in 
Eq.~(\ref{matter}) 
by a chiral superfield ${\cal M}$. The expectation 
value of the lowest component,
$$
\langle {\cal M} \rangle = m\, ,
$$
yields the supersymmetric matter mass term.
The expectation value of the $F$ component generates the squark 
(selectron) mass. If
\beq
{\cal M} = m (1- b\theta^2)
\eeq
where $b$ is a parameter of dimension of mass, the 
non-supersymmetric squark (selectron) mass term takes the form
\beq
\Delta {\cal L}_m = - mb\phi\tilde\phi + \mbox{H.c.} \, 
\label{NONSUSY}
\eeq
where $\phi$ and $\tilde\phi$ are the lowest components
of the superfields $Q$ and $\tilde Q$. 

In order to use the holomorphic nature of the gauge kinetic term and 
the superpotential
we introduce an infrared cut-off parameter $\mu$
and the ultraviolet cut-off parameter $\Lambda$. It is assumed that
$\mu \gg m, b, m_{\tilde g}$, while $\Lambda$
is much larger than any of the physical parameters of dimension of 
mass. In principle, the ultraviolet cut-off
parameter $\Lambda$ can be regarded as a chiral superfield, too.
The theory is regularized in the ultraviolet by introducing the
Pauli-Villars fields (within the background field technique),
and higher derivatives. We do not need to know the precise form of 
the regulator sector. All we need to know is that such a 
regularization exists, and it preserves supersymmetry. The 
ultraviolet parameter $\Lambda$ is the mass of the Pauli-Villars 
fields
or a dumping factor in the covariant derivative term.
If $\Lambda$ is treated as a chiral superfield, we assume that only 
its 
lowest component develops an expectation value.

In evolving the Lagrangian~(\ref{matter}) from the ultraviolet point 
$\Lambda$ down to $\mu$ a $Z$ factor appears in front of  the 
kinetic 
term of the matter fields,
$$
\left( \bar Q  e^V Q + \bar{\tilde Q}  e^{-V} \tilde Q \right) 
\ra Z \left( \bar Q  e^V Q + \bar{\tilde Q}  e^{-V} \tilde Q \right) \, .
$$
In the theory where the gauge coupling is substituted by ${\cal S}$
the $Z$ factor becomes a superfield, too. We will denote this 
superfield by ${\cal Z}$; its decomposition takes the form
\beq
{\cal Z} = Z\left( 1 + \frac12 \zeta\theta^2 + \frac12 \zeta^\dagger 
\bar\theta^2 + ... 
\right)\, .
\label{dzeta}
\eeq
Other components than those indicated above are irrelevant for our 
consideration; $\zeta$ and $\zeta^\dagger$ must be (and actually 
are) treated
 in the linear approximation.
Then, it is convenient to introduce
\beq
{\cal Z}_L =  Z\left( 1 + \zeta\theta^2 \right)
\eeq
and
\beq
{\cal Z}_R =  Z\left( 1 + \zeta^\dagger \bar\theta^2 \right)\, .
\eeq

\section{Supersymmetric Electrodynamics}

We begin our derivations from SQED since the relation between the 
gauge couplings in $\Gamma$ and $S$ are especially simple in this 
case. Let us first assume that in the bare Lagrangian
$b_0$ is put to zero, so that the only source of SUSY breaking is the
photino mass. We will see that in evolving the theory from 
$\Lambda$ 
down to $\mu$ we do generate the selectron mass $b$, necessarily.

Let us briefly remind how the exact $\beta$ function is obtained in
SQED without SUSY breaking \cite{SVZ,ShiVa}. The relation between 
the Wilsonean gauge coupling and that in $\Gamma$ is
\beq
\left( \frac{8\pi^2}{g^2}\right)_W =
\frac{8\pi^2}{g^2} + 2\ln Z\, ,
\label{wnw}
\eeq
where $Z$ stands for  the $Z$ factor of the matter fields, and the
renormalization of the Wilsonean gauge coupling is exactly one-loop,
\beq
\left( \frac{8\pi^2}{g^2}\right)_W = \left( 
\frac{8\pi^2}{g^2_0}\right)_W + 2 \ln\frac{\Lambda}{\mu}\, .
\label{wcr}
\eeq
Thus,  the Gell-Mann-Low function for the Wilsonean couplings is 
one-loop. The conventional definition of the coupling constants 
refers, however, to $\Gamma$, not to the Wilsonean action. 
Then combining Eqs.~(\ref{wnw}) and (\ref{wcr})  we arrive at
\cite{SVZ}
\beq
\beta (\alpha ) =\frac{\alpha^2}{\pi }\left[
1-\gamma (\alpha ) \right]
\, ,
\label{QEDbeta}
\eeq
where $\gamma (\alpha ) $ is the anomalous dimension of the 
electron (selectron) field,
\beq
\gamma =-\mu \frac{d\ln Z}{d \mu}\, .
\label{gamma}
\eeq
In the leading (one-loop) order
$$
Z = 1 
-\frac{\alpha}{\pi}\ln\frac{\Lambda}{\mu}\,\,\,\mbox{and}\,\,\, 
\gamma = -\frac{\alpha}{\pi }\, .
$$

What is to be changed in this derivation if $1/g^2$ is substituted by
a superfield ${\cal S}$? 
It is clear that the one-loop nature of the renormalization of the
gauge kinetic term in the 
Wilsonean action, Eq.~(\ref{wcr}), remains intact. The only difference 
is the
fact that $Z\ra {\cal Z}$, and ${\cal Z}$ depends now on ${\cal S}$. An 
additional term in ${\cal Z}$
arises, linear in $F_{\cal S}$. This term is obviously involved in the
renormalization of the photino mass. An analog of Eq.~(\ref{wnw}), 
describing the transition from the Wilsonean gauge coupling to that
in $\Gamma$, now  takes the form \footnote{
When writing the action in terms of the renormalized fields $Q_r$ 
and $\tilde Q_r$,
\begin{eqnarray}
Q_r (\tilde Q_r)  &=& {\cal Z}_L^{1/2} Q ({\cal Z}_L^{1/2} \tilde Q) \, ,
\nonumber\\
\bar{Q} (\bar{\tilde Q_r}) &=& {\cal Z}_R^{1/2} \bar{Q} ({\cal 
Z}_R^{1/2} \bar{\tilde Q}) \, , 
\nonumber
\end{eqnarray}
the Konishi anomaly \cite{Koni1} generates terms
$$
\frac1{32 \pi^2} \int d^2 \theta (\ln{{\cal Z}_L}) W^2 +{\rm H.c.}
$$
as a Jacobian of the measure \cite{Koni2}. Thus, Eq.~(\ref{wnwsb}) 
is justified.
}
\beq
(8\pi^2 {\cal S})_W = (8\pi^2 {\cal S}) + 2\ln {\cal Z}_L\, ,
\label{wnwsb}
\eeq
and
\beq
(8\pi^2 {\cal S}(\mu ))_W = (8\pi^2 {\cal S}_0)_W + 2\ln 
\frac{\Lambda}{\mu}\, .
\label{wcrsb}
\eeq
Since there is no $F$ component in the logarithm of $\Lambda/\mu$
and ${\cal Z}_0$ is put to unity, by definition, we conclude that
\beq
\left[(4\pi^2 F_{\cal S}) +  \zeta \right]_\mu
=\left[(4\pi^2 F_{\cal S}) \right]_{\Lambda}\, .
\label{defrel}
\eeq

Let us discuss  the dependence of the ${\cal Z}$ factor
on the superfield ${\cal S}$. 
As a warm-up exercise consider the leading-log approximation for 
the $Z$ factor.  The corresponding analysis will determine
the running of $m_{\tilde g}$ up to two loops.
In the leading-log approximation \footnote{
Here it is necessary  to note that $\alpha$ in the $Z$ 
factor means 
$1/\left(2\pi ({\cal S}+{\cal S}^{\dagger})\right)$
after $1/g^2$ is substituted by ${\cal S}$.
}
\beq
Z =\frac{\alpha (\mu )}{\alpha_0} \,\, \ra \,\,  
{\cal Z}_L = \frac{\alpha (\mu )}{\alpha_0}\frac{1-2m_{\tilde g 
0}\theta^2}{1-2m_{\tilde g}\theta^2}\, .
\eeq
In other words,
\beq
\zeta = 2 \left(m_{\tilde g} - m_{\tilde g 0} \right)\, .
\label{zlast}
\eeq
Using the definition of ${\cal S}$, see Eq.~(\ref{gmg}), and 
Eqs.~(\ref{defrel}) and  (\ref{zlast}) we immediately conclude that
\beq
\frac{m_{\tilde g}}{\alpha}\left( 1-\frac{\alpha}{\pi} \right)
= \mbox{RGI}\, . 
\eeq
This relation is nothing but the two-loop truncation of the general 
expression~(\ref{RGI}).

Generically
\beq
Z = \exp\int_\alpha^{\alpha_0}\frac{\gamma (\alpha )}{\beta(\alpha 
)} d\alpha\, .
\label{zgen}
\eeq
Inclusion of the $F$ component of the superfield ${\cal S}$
reduces to the following changes in Eq.~(\ref{zgen}):
$$
\alpha_0 \ra \alpha_0 + 2m_{\tilde g 0}\alpha_0\,  \theta^2\, , \,\,\,
\alpha \ra \alpha + 2m_{\tilde g }\alpha \, \theta^2\, , \,\,\, Z \ra 
{\cal Z}_L\, .
$$
Therefore, the all-order result for $\zeta$ is
\beq
\zeta = 2\left[ -\frac{\alpha \gamma (\alpha )}{\beta (\alpha )} 
m_{\tilde g}  + \frac{\alpha_0 \gamma (\alpha_0 )}{\beta (\alpha_0 
)} m_{\tilde g 0 }\right] \, .
\label{zetagen}
\eeq
Using this result in Eq.~(\ref{defrel}) we obtain the all-order
prediction for the running  of the photino mass presented in the 
general relation~(\ref{RGI}).  This relation is valid as long  as there 
are 
no trilinear couplings in the superpotential.  

Note that even though at the ultraviolet cut-off  the mass parameter
was assumed above to be supersymmetric, i.e. $b_0 =0$, the 
evolution from $\Lambda$ down to $\mu$ does produce a 
non-supersymmetric selectron mass. Since ${\cal M} = m_0/{\cal 
Z}_L$, it is 
not difficult to see that
\beq
b = \zeta = 2\pi \left\{ 
- \frac{m_{\tilde g}}{\alpha}\frac{\gamma (\alpha )}{1- \gamma 
(\alpha ) } +
\frac{m_{\tilde g 0}}{\alpha_0}\frac{\gamma (\alpha_0 )}{1- 
\gamma (\alpha_0 ) }
\right\} \, .
\label{bz}
\eeq
Therefore, Eq.~(\ref{defrel}) can be obviously rewritten as
\beq
\frac{m_{\tilde g}}{\alpha} - \frac{b}{2\pi} = \mbox{RGI}\, .
\label{mb}
\eeq
In this form the prediction is valid even if a non-vanishing
selectron mass $b_0\neq 0$ is introduced in the original Lagrangian,
as long as $\mu \gg m, b$. Also,  even if there is a trilinear coupling 
in the superpotential, this is valid. This assertion follows from the 
fact that 
Eq.~(\ref{mb}) can be proven directly  from  Eq.~(\ref{defrel}),
being combined with the   relation between $b$ and $b_0$
\begin{equation}
b = b_0 + \zeta\, . 
\end{equation}
The photino mass, the selectron mass 
and the gauge coupling constant run in such a way that the 
combination~(\ref{mb}) 
stays $\mu$ independent. 

So far the normalization point $\mu$ was assumed to lie above the 
mass thresholds. 
In conclusion of this section we turn to the question what happens if 
the
evolution of the gauge coupling is complete, i.e. the normalization 
point $\mu $ becomes  lower than $m$ -- we run all the way down 
till the point where the gauge coupling becomes frozen. Likewise, the 
evolution
of $m_{\tilde g}$  freezes at $\mu /m \ra 0$. (It is assumed that
$m_{\tilde g} \ll m$.)
As was noted in  Ref.~\cite{RR},
some curious relations between the frozen low-energy values of the 
parameters and those in the original Lagrangian emerge in this 
formulation. 
If we dive below the threshold of the matter fields the  exact 
expression for the gauge coupling constant 
looks 
as if it were exactly one-loop, but with the fake value of the 
threshold,
\beq
\alpha_{\rm LE} = 
{\alpha_0}\left\{ 
1+\frac{\alpha_0}{\pi}\ln\frac{\Lambda}{m_0}\right\}^{-1}\, .
\label{ale}
\eeq
Here the subscript LE marks the low-energy (frozen) quantities.
This result is known for a long time \cite{SVZ}. We can get a similar 
expression for the photino  mass. Passing through the matter 
threshold modifies Eq.~(\ref{mb}).
Say, if we descend down to the domain of freezing 
\beq
\left( \frac{m_{\tilde g}}{\alpha} 
\right)_{\rm LE} = \frac{m_{\tilde g 0}}{\alpha_0} - 
\frac{b_0}{2\pi}\, . 
\label{mle}
\eeq
The second term on the right-hand side of Eq.~(\ref{mle}) 
corresponds to a finite correction to the gaugino mass,  Fig. 1.
There is no explicit $\gamma$ factor here; all non-triviality
associated with $\gamma$ is hidden completely. 

It is worth emphasizing that Eqs.~(\ref{ale}) and (\ref{mle})
take into account the  threshold at $\mu \sim m$
in full and exactly.

   \begin{center} \begin{picture}(300,100)(0,0)
  \ArrowArc(150,50)(35,0,180) \Text(175,50)[c]{$Q$}
   \ArrowArc(150,50)(35,180,0) \Text(128,50)[c]{$\tilde{Q}$}
   \Photon(185,50)(205,50){4}{4}  \Text(215,50)[c]{$\lambda$}
  \Photon(115,50)( 95,50){4}{4}  \Text( 85,50)[c]{$\lambda$}
   \Vertex(150,15){3}           \Text(150, 5)[c]{$m_0$}
   \Vertex(150,85){3}           \Text(150,95)[c]{$m_0 b_0 \theta^2$}
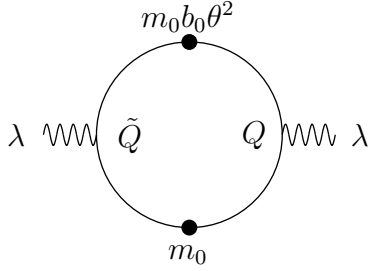
   \end{picture}  \\ {Fig. 1. The contribution in the gaugino mass
     arising from SUSY violating selectron mass.} \end{center}

\section{Supersymmetric Gluodynamics}

Our task now is to extend the method to non-Abelian theories.
Although technically this case is somewhat more complicated, 
conceptually we will encounter no new elements. Let us first treat 
the case when there are no matter fields.

The main distinction is that even in the absence of the
mater fields the Wilsonean couplings are now different from those in 
$\Gamma$. 
According to Eq.~(\ref{gmg}), the Wilsonean coupling
\beq
\left( \frac{1}{ g_0^2} \right )_W \ra \left(\frac{1 - 2 m_{\tilde 
g}\theta^2}{g^2}\right )_W\, , 
\label{nasubs}
\eeq
where $m_{\tilde g W}$ is a Wilsonean gluino mass. Below we will 
need to exploit the 
fact
that $m_{\tilde g W}$ is {\em not} what is usually called the gluino 
mass. Indeed, the mass term is usually defined as a parameter in 
front of
$(1/g^2)\lambda\lambda$
in $\Gamma$.
It is not difficult to establish a relation between these two 
parameters. If in the Wilsonean Lagrangian the mass perturbation is
\beq
\left( \frac{m_{\tilde g}}{ g^2} \right)_W \lambda\lambda\, ,
\eeq
in passing to $\Gamma$, we get
\beq
\left( \frac{m_{\tilde g}}{ g^2} \right)_W  \frac{1}{1 - 
(T(G) g^2)/(8\pi^2)} 
\lambda\lambda |_{ext}\, ,
\label{passing}
\eeq
to be identified with
$$
\frac{m_{\tilde g}}{ g^2}\, .
$$
From here we conclude that
\beq
\frac{m_{\tilde g}}{ g^2} \left (1 - \frac{T(G)g^2}{8\pi^2}\right) = 
\left( 
\frac{m_{\tilde g}}{ 
g^2} \right)_W \, .
\label{WM}
\eeq
Note that in obtaining Eq.~(\ref{passing}) we used the fact 
\cite{ShiVa} that
the matrix element of the operator $W^2$ is
$$
\frac{1}{1 - (T(G)g^2)/(8\pi^2)} W^2|_{ext}\, .
$$

Once the relation between the Wilsonean and conventional
mass parameters is established, we can forget for a while about
the conventional parameter, and focus on what happens with the 
Wilsonean action under renormalizations. As known from 
Ref.~\cite{Shif}, 
in the Wilsonean action, where the infrared 
contributions are not included by definition, the holomorphy is 
preserved 
-- $F$ terms should depend only on the chiral superfields, not 
antichiral, and so on.
In the Wilsonean action the coupling constant is renormalized only 
at one loop,
\beq
\left( \frac{1}{ g^2} \right)_W = \left( \frac{1}{ g_0^2} \right)_W -
\frac{3T(G)}{8\pi^2}\ln \frac{\Lambda}{\mu}\, .
\label{E1}
\eeq
Now, in this relation it is perfectly legitimate  to substitute
each  bracket by the same bracket with $(1 - 2 m_{\tilde 
g}\theta^2)$
in the numerator.
Both, the left- and right-hand sides appear as coefficients of the $F$ 
terms in the Wilsonean action, and we multiply by 
the chiral superfield. From Eq.~(\ref{E1}), inspecting the coefficient in 
front of $\theta^2$,  we immediately deduce
that
\beq
\left( \frac{m_{\tilde g}}{ g^2} \right)_W = \mbox{RGI}\, .
\label{E2}
\eeq
Invoking now Eq.~(\ref{WM}) we see
that 
\beq
\frac{m_{\tilde g}}{ g^2} \left (1 - \frac{T(G) g^2}{8\pi^2}\right) = 
\mbox{RGI}\, .
\label{E3}
\eeq
Taking into account the explicit form of the NSVZ $\beta$ function in 
the case at hand, see Eq.~(\ref{GMLF}),  we conclude that 
Eq.~(\ref{E3}) is
in full accord with the general expression~(\ref{RGI}). 

The same result  can be obtained in a slightly different way.
The vacuum expectation value of the {\em operator}
$W^2$ is a physical quantity; it is  RG invariant. 
This vacuum expectation value can be written as \cite{Shif3,Shif}
\beq
\langle W^2\rangle = (\mbox{Numer.Const.})\times \mu^3 \exp \left( 
-4\pi^2
{\cal S}_W\right) \, . 
\label{E4}
\eeq
In both the left and right-hand sides we have only chiral
superfields, as it should be. If ${\cal S}$ develops
an $F$ term, we expand in it. The expectation value of the $F$ term
of the operator $W^2$ is proportional to the vacuum energy density
\footnote{This is another reason for the absence of renormalization.
 }. The $F$ term on the right-hand side is proportional to
$(m_{\tilde g}/g^2)_W$. In this way we arrive at Eq.~(\ref{E2}). 

If the matter fields are switched on, the argument becomes 
somewhat  
more subtle in the part referring to the matter field $Z$ factors.  Let 
us sketch here the basic points.  The main idea is to continue using 
the Wilsonean action.

Under the renormalization the Wilsonean action
$$
\left( \frac{1-2m_{\tilde g 0}\theta^2}{ g_0^2} \right)_W W^2
$$
goes into 
\beq
\left\{ \left( \frac{1-2m_{\tilde g0}\theta^2}{ g_0^2} \right)_W 
+\left[
\frac{\sum_i T(R_i)-3T(G)}{8\pi^2}
\ln \frac{\Lambda}{\mu} - \sum_i\frac{ T(R_i)}{8\pi^2}
\ln {\cal Z}_{Li}\right]
\right\} W^2\,  ,
\eeq
where $T(R_i)$ are the Dynkin indices for the
matter fields. In the fundamental representation $T=1/2$
for each subflavor;  $T=1$ for one flavor. The sum runs over all 
matter fields. 

We must now derive an  analog of Eq.~(\ref{zetagen}).
A straightforward calculation yields
$$
\zeta_i = 4\pi \left\{ \left( \frac{m_{\tilde g}}{\alpha }\right)_W 
\frac{\gamma_i}{3T(G) -\sum_i T(R_i)(1-\gamma_i)}\right.
-
$$
\beq
\left.
\left( \frac{m_{\tilde g 0}}{\alpha_0 }\right)_W 
\frac{\gamma_{0i}}{3T(G) -\sum_i T(R_i)(1-\gamma_{i0})}
\right\}.
\label{E5}
\eeq
Here $\gamma_i = \gamma_i(\alpha )$ and $\gamma_{i0} = 
\gamma_i(\alpha_0 )$. From  where we immediately conclude that
\beq
\left( \frac{m_g}{ g^2} \right)_W \left(
1-\frac{T(R_i)\gamma_i}{3T(G)-T(R_i)(1-\gamma_i)}\right)
=\mbox{RGI}\, .
\label{E6}
\eeq
Invoking again Eq.~(\ref{WM})
and using the fact that 
the NSVZ $\beta$ function in the case at hand has the form
\beq
\beta(\alpha ) = -\frac{\alpha^2}{2\pi}\, 
\frac{3T(G)-T(R_i)(1-\gamma_i)}{1- T(G)\alpha (2\pi )^{-1}} \, ,
\label{NSVZbeta}
\eeq
 we reproduce Eq.~(\ref{RGI}). \footnote{The anomalous dimensions 
of the matter fields {\em per se} cannot be determined to all
orders since the holomorphy arguments do not apply in this case.
At  one loop  $\gamma_i = - C(R_i)\alpha /\pi$ where
$C(R_i) =T(R_i)$ dim(adj)/dim($R_i$).}

An alternative form of the RGI relation is obtained
if the squark mass terms are introduced. As in SQED,
$b_i = b_{i0} +\zeta_i$, which implies, in turn, that
\beq
\frac{m_{\tilde g}}{\alpha}
\left( 1-\frac{T(G)\alpha}{2\pi}
\right) - \sum_i \frac{T(R_i)b_i}{4\pi}
=\mbox{RGI}\, ,
\label{mb2}
\eeq
to be compared with Eq.~(\ref{mb}) in SQED. 

\vspace{0.3cm}

\section{Exact GUT Relation of the Gaguino Masses}

So far, we assumed that the chiral multiplets, $Q$ and $\tilde{Q}$, do 
not
have vacuum expectation values, that is, the spontaneous breaking of 
the gauge 
symmetry  does not occur.  It is interesting to discuss the
issue of the gaugino mass renormalization in the presence of the
spontaneous breaking of the gauge symmetry, keeping in mind
possible applications in theories of Grand Unification (GUT). 

Let us consider $SU(3)$ gauge model as a prototype  example;
we will introduce a chiral multiplet $\Sigma^a$ in the  adjoint 
representation ($a = 1,2, ... , 8$). The vacuum expectation value of 
$\Sigma$ 
 induces the gauge symmetry breaking, $SU(3)\rightarrow 
SU(2)\times U(1)$ (see Ref.~\cite{RR} for the discussion of the 
running gauge coupling in this model). 

The original superpotential of the adjoint chiral multiplet is 
\beq
P
= 
\frac1{4g_0^2}
\left(m_0 \Sigma^a \Sigma^a + y_0 d_{abc} 
\Sigma^a\Sigma^b\Sigma^c \right)
\, ,
\label{spsu3}
\eeq
where $d_{abc}$ are the $d$ symbols of $SU(3)$. 
Here again, we will substitute the mass $m$ and the Yukawa 
coupling $y$ by the chiral superfields ${\cal M}$ and ${\cal 
Y}$, respectively, and 
assume 
non-vanishing vacuum expectation values of the $F$ components, in 
order 
to introduce  soft SUSY breaking,
\begin{eqnarray}
\langle {\cal M}_0 \rangle &=& m_0 (1 - b_0 \theta^2) \, , 
\nonumber\\
\langle {\cal Y}_0 \rangle &=& y_0 (1 - a_0 \theta^2) \, .
\end{eqnarray}
Here $a$ and $b$ are parameters of dimension of mass. We assume 
that
$a$, $b$, and $m_{\tilde{g}}$ $\ll m$.
By inspecting the superpotential (\ref{spsu3}) we observe that 
$\Sigma$ gets the  following vacuum expectation 
value 
\beq
\langle\Sigma \rangle = 
\frac{1}{2\sqrt{3}}
\left(
\begin{array}{ccc} 
1&&\\
&1&\\
&&-2
\end{array}
\right) {\cal V}_0  
\eeq
breaking $SU(3)$ down to $SU(2)\times U(1)$;
here \footnote{
In the diagrammatic calculation of the radiative correction to the 
gaugino masses 
in this model, we have to keep the  shift of the scalar component of 
${\cal V}$ from its supersymmetric value due to introduction of $a$ 
and $b$.
This is because this shift leads to a non-vanishing mass of the 
fermionic 
partners 
of the Nambu-Goldstone bosons \cite{HMG}. However, the result for 
the gaugino mass is not changed compared to  ours. This follows from 
the 
fact 
that the shift of the scalar component of ${\cal V}_0$ has a 
correlation 
with 
the value of the $F$ component of ${\cal V}_0$.
}
\begin{eqnarray}
{\cal V}_0 &=& \frac{2 {\cal M}_0}{\sqrt{3}{\cal Y}_0}\left(1 + 
O(b/m, a/m)\right)
\nonumber\\
&=&  \frac{2 m_0}{\sqrt{3}y_0} (1 + (a_0-b_0) \theta^2).
\eeq
Due to the vacuum expectation value ${\cal V}_0$ four out of eight 
gauge multiplets 
get a 
mass and form, together with  two $SU(2)$ doublets from  
$\Sigma$ eaten up in the super-Higgs mechanism, massive
 vector multiplets. One $SU(2)$ triplet and one singlet from  
$\Sigma$ 
survive.
The mass of the $SU(2)$ triplet is 
\beq
M_{\Sigma} = {\cal M}_0\, .
\eeq 

Below the masses of the heavy gauge bosons (the ``elephants")
and the masses of the surviving fields from $\Sigma$ (i.e. below the 
unification thresholds) the $SU(2)$ and $U(1)$ gauge couplings 
evolve separately, and so do the corresponding gaugino masses.
The gauge couplings diverge. Our task is to express the low-energy 
values
of the gauge couplings and gaugino masses (far below the thresholds)
in terms of the high-energy (i.e. above-threshold) parameters.

In other words, we choose the normalization point $\mu$ far below 
the lowest 
threshold
and the ultraviolet cut-off $\Lambda$ far above the highest one.
The Wilsonean couplings of the $SU(2)$ and $U(1)$ gauge multiplets 
are given by
\begin{eqnarray}
\left({\cal S}_{SU(2)}\right)_W &=&
 \left({\cal S}_0\right)_W
-  \frac{6}{8\pi^2} \ln \frac{\Lambda}{\mu}
-  \frac{2}{8\pi^2} \ln \frac{\Lambda}{\sqrt{3}{\cal V}_0/2}
+  \frac{2}{8\pi^2} \ln \frac{\Lambda}{{\cal M}_0}\, ,
\label{SU(2)}
\\
\left({\cal S}_{U(1)}\right)_W &=&
 \left({\cal S}_{0}\right)_W 
-  \frac{6}{8\pi^2} \ln \frac{\Lambda}{\sqrt{3}{\cal V}_0/2}\, .
\label{U(1)}
\end{eqnarray}
The second and fourth terms in Eq.~(\ref{SU(2)}) 
are the contributions from the $SU(2)$ gauge multiplet and 
SU(2) triplet in $\Sigma$, respectively, and the others come from 
the massive vector multiplets. By taking the $F$ components of 
Eqs.~(\ref{SU(2)}) and (\ref{U(1)}) and using Eq.~(\ref{WM})
we arrive at
\begin{equation}
\frac{m_{\tilde{g}_2}}{\alpha_2} \
\left (1 - \frac{T(SU(2)) \alpha_2}{2\pi} \right) 
=
\frac{m_{\tilde{g}0}}{\alpha_0} 
\left (1 - \frac{T(SU(3)) \alpha_0}{2\pi} \right) 
-  \frac{1}{2\pi} (a_0-b_0)
-  \frac{1}{2\pi}  b_0\, ,
\label{SU(2)n}
\end{equation}
\begin{equation}
\frac{m_{\tilde{g}_1}}{\alpha_1} 
=
\frac{m_{\tilde{g}0}}{\alpha_0} 
\left (1 - \frac{T(SU(3))\alpha_0}{2\pi}\right) 
-  \frac{3}{2\pi} (a_0-b_0)\, .
\label{U(1)n}
\end{equation}
Here $m_{\tilde{g}_2}$ ($m_{\tilde{g}_1}$) and $\alpha_2$ 
($\alpha_1$)
are the $SU(2)$ ($U(1)$) gaugino mass and gauge coupling constant
at $\mu$. \footnote
{
It can be proven by using the relation 
$$
\mu\frac{d a}{d\mu} = \frac32 \mu\frac{d b}{d\mu} \, .
$$
that above 
the threshold these equations give the same RG expressions.
}
Then, we can get an exact relation for the gaugino mass at low 
energy
\begin{equation}
\frac{m_{\tilde{g}_2}}{\alpha_2} 
\left(1- \frac{T(SU(2))\alpha_2}{2\pi}\right) 
-
\frac{m_{\tilde{g}_1}}{\alpha_1} 
= 
\frac{1}{\pi} (a_0 - b_0) - \frac{1}{2\pi} b_0 .
\label{diff}
\end{equation}
Equation (\ref{diff}) is usually referred to as the  {\em GUT relation}.
It is worth emphasizing that the relation we derived is valid
to all orders in the coupling constants, and exactly takes into account
the threshold effects.

In the diagrammatic calculation, the first term on the right-hand side 
in Eq.~(\ref{diff}) comes from the mass of the fermionic partners of 
the Goldstone bosons, and the second term comes from a diagram 
similar to that of
Fig. 1,  in which the surviving $SU(2)$ triplet from $\Sigma$ 
propagates 
\cite{HMG}.

Let us consider a particular case $a_0 = b_0 = 0$, i.e. all
 SUSY breaking parameters, except  the gaugino mass, 
are zero at $\Lambda$. 
It is remarkable  that in this case the  ratio of the low-energy
gaugino masses 
is completely determined by {\em only} the low-energy gauge 
couplings, with no  
dependence on details of the model at the gauge symmetry breaking 
scale,
$$
\frac{m_{\tilde{g}_2}}{m_{\tilde{g}_1}}
=\frac{\alpha_2}{\alpha_1}\left( 1-\frac{\alpha_2}{\pi}\right)^{-1}\, 
.
$$

\section{Confronting our Results with Explicit Calculations at Two-
Loop Level}

Equation~(\ref{RGI}) is our basic all-order prediction for the running 
gaugino mass. It is instructive to check it 
``empirically". The Gell-Mann-Low function is scheme-independent 
up to two loops. The $\gamma$ factors are scheme-independent only 
in the leading (one-loop) order, but, as we have seen, the one-loop  
$\gamma$ factors will affect the renormalization of $m_{\tilde g}$
at two loop level only. Therefore, the running of $m_{\tilde g}$ up to 
two loops 
is unambiguously given by Eq.~(\ref{RGI}) if there are no trilinear 
couplings 
in the superpotential.  Here we compare
our result with the direct two-loop calculation of the gluino mass
reported in Ref.~\cite{yamada} where the $\overline{DR}$ scheme is 
adopted. 

The two-loop RG equations of the gauge coupling constant and the 
gaugino mass 
in the $\overline{DR}$ scheme are given as follows,
\begin{eqnarray}
\mu\frac{d}{d \mu} \alpha &=&
  \frac{\alpha^2}{2\pi} \left\{\sum_iT(R_i) -3 T(G)\right\}
\nonumber\\
&&
+ \frac{\alpha^2}{2\pi} \left\{
\left(\sum_iT(R_i) -3 T(G) \right) T(G) \frac{\alpha}{2\pi} 
- \sum_i T(R_i) \gamma_i \right\}
\label{gauge}
\end{eqnarray}
\begin{eqnarray}
\mu \frac{d}{d \mu} \left(\frac{m_{\tilde{g}}}{\alpha} \right) &=&
  \frac{1}{2\pi} \left\{
\left(\sum_iT(R_i) -3 T(G) \right) T(G) \frac{\alpha}{2\pi} 
- \sum_i T(R_i) {\tilde{\gamma}}_i \right\} \, .
\label{gaugino}
\end{eqnarray}
Here $\gamma_i$ and $\tilde{\gamma}_i$ are defined as
\begin{eqnarray}
\gamma_i &=& - \mu \frac{d \ln Z_i}{d \mu} \, ,\nonumber\\
\tilde{\gamma}_i &=& - \frac12 \mu \frac{d \zeta_i}{d \mu} \, 
,\nonumber
\end{eqnarray}
and  are given, at the one-loop level,  by the following expressions
\begin{eqnarray}
\gamma_i &=& - \frac{\alpha}{\pi} C(R_i) \, , \nonumber\\
\tilde{\gamma}_i &=& - \frac{\alpha}{\pi} C(R_i)~ m_{\tilde{g}} \, . 
\label{one-loop}
\end{eqnarray}
(Note that $C$ is defined as $C \delta^i_j 
= \left(\sum_aT^a T^a\right)_i^{j}$, and $C=(N^2-1)/2N$ for the 
fundamental
representation of $SU(N)$.)
From Eqs.~(\ref{gauge}) and (\ref{gaugino}) we can get
\begin{eqnarray}
\mu\frac{d}{d \mu} \left(\frac{\alpha m_{\tilde{g}}}{\beta(\alpha)} 
\right) 
&=&
\frac{\alpha^3}{[\beta(\alpha )]^2} \frac{1}{2\pi} 
\sum_i T(R_i) \left\{\beta(\gamma_i) m_{\tilde{g}}- 
\frac{\alpha}{2\pi} 
(\sum_iT(R_i) -3T(G)) \tilde{\gamma}_i \right\} \, ,
\nonumber\\
\label{check}
\end{eqnarray}
where $\beta(\gamma_i)$ is the $ \beta$ function for $\gamma_i$,
$$
\beta(\gamma_i) = \mu \frac{d \gamma_i(\alpha )}{d \mu}\, . 
$$
It is easy to  prove,  by using the explicit forms of $\gamma_i$ and
$\tilde{\gamma}_i$ at one-loop level, Eqs.~(\ref{one-loop}), that the 
right-hand side  of 
Eq.~(\ref{check}) vanishes. Thus, Eq.~(\ref{RGI}) is confirmed
at two-loop level. 

Equations (\ref{mb}) and (\ref{mb2}) can be derived  directly by 
using the 
RG 
equation for 
the SUSY breaking parameter $b_i$,
\begin{equation}
\mu \frac{d b_i}{d \mu}= - 2 \tilde{\gamma}_i \, .
\end{equation}
Here explicit forms of $\gamma$ and $\tilde{\gamma}$ are not 
needed,
as is expected from the fact that Eqs.~(\ref{mb}) and (\ref{mb2}) are 
valid even 
in the presence of the  
 trilinear couplings in the superpotential.

\section{Conclusions and Discussion}

In this paper, in SUSY gauge theories with the soft SUSY 
breaking, we studied the running of the gaugino and/or squark 
(selectron) masses. 
Several exact (i.e. all-order) predictions are obtained
by 
exploiting the  relation between the gaugino masses in the generator 
of the 
1PI vertices and the Wilsonean action and by 
using the 
holomorphic nature of the $F$ term in the Wilsonean action. 
Our main findings can be summarized as follows.

\begin{itemize}
\item In the absence of the trilinear couplings in the
superpotential we derived a general {\em exact} formula
relating the running gluino mass to the running gauge coupling 
constant,
\beq
\frac{\alpha m_{\tilde g}}{\beta (\alpha )} = \mbox{RGI}\, ,
\eeq
This formula is  valid even if the matter sector is chiral, 
no mass terms are possible (e.g. $SU(5)$ theory with an equal 
number of quintets and antidecuplets). 

\item If the matter sector is non-chiral, i.e.  supersymmetric 
mass terms are possible, and the squark SUSY-breaking masses are 
introduced  trough Eq. (\ref{NONSUSY}), then 
\beq
\frac{m_{\tilde g}}{\alpha}
\left( 1-\frac{T(G)\alpha}{2\pi}
\right) - \sum_i \frac{T(R_i)b_i}{4\pi}
=\mbox{RGI}\, .
\eeq
This formula is  valid even if there are trilinear couplings in the 
superpotential.

\item In certain instances the mass threshold effects in the
gaugino mass can be taken into account exactly. In SQED this 
assertion is illustrated by Eq. (\ref{mle}), in non-Abelian GUT's
by Eq. (\ref{diff}).

\end{itemize}

\vspace{0.3cm}

As was just mentioned, if the Yukawa (trilinear)
couplings in the superpotential are present (let us call them 
generically $h$) 
the exact result that we managed to get refers
to a linear combination of the gluino and (SUSY breaking) squark 
mass terms, rather than to the gluino mass  {\em per se}. 
Technically the reason is evident: unlike the case of pure gauge 
interactions now the derivative of $Z(\alpha, h)$ with respect to 
$\alpha$ is unrelated with $\gamma (\alpha ,h)$. It is still possible 
to obtain the exact expressions of the type (\ref{RGI}), in the closed 
form, for a specific choice of the points on the $\{\alpha , h\}$ plane.
We mean fixed points of 
the gauge coupling constant and the Yukawa coupling constants. If 
the initial set of parameters is such that the condition 
\begin{equation} 
\gamma (\alpha , h) =\frac{1}{3} \frac{\beta(\alpha)}{\alpha}\, ,
\label{fixedpoint}
\end{equation}
is met,  
the ratio of  the gauge coupling constant to the Yukawa 
coupling constant in the superpotential is  RG invariant. 
In Refs.~\cite{fixed} it was shown,
by examining  explicit formulae of the RG equations,  that 
 Eq.~(\ref{fixedpoint}) does have a solution at least up  two loops 
(the so called ``$P=Q/3$'' rule). More interestingly, the authors of  
Refs.~\cite{fixed} argue that
if Eq.~(\ref{fixedpoint}) is satisfied, the gaugino mass and the 
SUSY breaking trilinear scalar coupling constant $a$, associated 
with the 
Yukawa coupling, have a fixed point at 
\begin{equation}
a=-m_{\tilde{g}}\, .
\label{a_gaugino} 
\end{equation}

Now we are able to solve the question of the  fixed-point behavior 
of the SUSY breaking parameters beyond two loops. Equation 
(\ref{a_gaugino})  follows from the holomorphic 
nature of the $F$ terms. In fact, it can be proven that 
Eq.~(\ref{a_gaugino}) 
is valid to all orders provided  Eq.~(\ref{fixedpoint}) is satisfied to all 
orders. If that's the case
 Eq.~(\ref{RGI}) remains valid even in the presence of the Yukawa 
couplings.

What remains to be done? 
So far, we have not considered models with the chiral
matter sector (no  mass term possible), with the Yukawa (trilinear) 
couplings of the 
chiral superfields in the superpotential. 
It is possible to show that a RGI relation for the 
gaugino mass {\it at any order} has the form
\beq
\frac{m_{\tilde g}}{\alpha}
\left( 1-\frac{T(G)\alpha}{2\pi}
\right) - \sum_i \frac{T(R_i)\zeta_i}{4\pi}
=\mbox{RGI}\, .
\eeq
The task is to evaluate
$\zeta_i$  from supersymmetric $Z$ factors 
of the
chiral multiplets. 

Another problem is quite obvious too. In Ref. \cite{JJ}
a perturbative renormalization $\overline{DR}$-related  scheme was 
identified
yielding the NSVZ $\beta$ function up to three loops.
The anomalous dimensions of the matter fields are known in this 
scheme in two loops. If the running of the gluino and squark masses
were known in three loops in this scheme one could have verified 
our all-order predictions by comparing them with the explicit 
calculations up to three loops. 

\vspace{1cm}

\underline{\bf Acknowledgments}: \hspace{0.2cm} 
 
\vspace{0.2cm}

One of the authors (J.H.)  would like to  thank H.~Murayama, 
K.~Okuyama,  
Y.~Yamada, and T.~Yanagida for useful discusssions. This work was 
supported in part by DOE 
under the grant number DE-FG02-94ER40823.

\end{document}